\documentclass[useAMS,usenatbib,a4paper]{mn2e}
\usepackage{amsmath}
\usepackage{amssymb}
\usepackage{amsfonts}
\usepackage {multirow}
\usepackage {graphicx}
\usepackage {natbib}
\usepackage{graphicx,amsmath}
\usepackage[hyperindex,breaklinks=true, colorlinks, citecolor=blue]{hyperref}  
\usepackage{enumerate}
\usepackage{tablefootnote}
\usepackage{eurosym}
\usepackage[usenames,dvipsnames]{xcolor}

\def\setsymbol#1#2{\expandafter\def\csname #1\endcsname{#2}}
\def\getsymbol#1{\csname #1\endcsname}

\def\Planck{\textit{Planck}}

\newbox\tablebox    \newdimen\tablewidth
\def\leaderfil{\leaders\hbox to 5pt{\hss.\hss}\hfil}
\def\tablenote#1 #2\par{\begingroup \parindent=0.8em
    \abovedisplayshortskip=0pt\belowdisplayshortskip=0pt
    \noindent
    $$\hss\vbox{\hsize\tablewidth \hangindent=\parindent \hangafter=1 \noindent
    \hbox to \parindent{$^#1$\hss}\strut#2\strut\par}\hss$$
    \endgroup}

\def\L2{\ifmmode L_2\else $L_2$\fi}

\def\DeltaT{\ifmmode \Delta T\else $\Delta T$\fi}
\def\deltat{\ifmmode \Delta t\else $\Delta t$\fi}
\def\fknee{\ifmmode f_{\rm knee}\else $f_{\rm knee}$\fi}
\def\Fmax{\ifmmode F_{\rm max}\else $F_{\rm max}$\fi}
\def\solar{\ifmmode{\rm M}_{\mathord\odot}\else${\rm M}_{\mathord\odot}$\fi}
\def\Msolar{\ifmmode{\rm M}_{\mathord\odot}\else${\rm M}_{\mathord\odot}$\fi}
\def\Lsolar{\ifmmode{\rm L}_{\mathord\odot}\else${\rm L}_{\mathord\odot}$\fi}
\def\inv{\ifmmode^{-1}\else$^{-1}$\fi}
\def\mo{\ifmmode^{-1}\else$^{-1}$\fi}
\def\sup#1{\ifmmode ^{\rm #1}\else $^{\rm #1}$\fi}
\def\expo#1{\ifmmode \times 10^{#1}\else $\times 10^{#1}$\fi}
\def\,{\thinspace}
\def\lsim{\mathrel{\raise .4ex\hbox{\rlap{$<$}\lower 1.2ex\hbox{$\sim$}}}}
\def\gsim{\mathrel{\raise .4ex\hbox{\rlap{$>$}\lower 1.2ex\hbox{$\sim$}}}}

\def\simprop{\mathrel{\raise .4ex\hbox{\rlap{$\propto$}\lower 1.2ex\hbox{$\sim$}}}}
\def\deg{\ifmmode^\circ\else$^\circ$\fi}
\def\pdeg{\ifmmode $\setbox0=\hbox{$^{\circ}$}\rlap{\hskip.11\wd0 .}$^{\circ}
          \else \setbox0=\hbox{$^{\circ}$}\rlap{\hskip.11\wd0 .}$^{\circ}$\fi}
\def\arcs{\ifmmode {^{\scriptstyle\prime\prime}}
          \else $^{\scriptstyle\prime\prime}$\fi}
\def\arcm{\ifmmode {^{\scriptstyle\prime}}
          \else $^{\scriptstyle\prime}$\fi}
\newdimen\sa  \newdimen\sb
\def\parcs{\sa=.07em \sb=.03em
     \ifmmode \hbox{\rlap{.}}^{\scriptstyle\prime\kern -\sb\prime}\hbox{\kern -\sa}
     \else \rlap{.}$^{\scriptstyle\prime\kern -\sb\prime}$\kern -\sa\fi}
\def\parcm{\sa=.08em \sb=.03em
     \ifmmode \hbox{\rlap{.}\kern\sa}^{\scriptstyle\prime}\hbox{\kern-\sb}
     \else \rlap{.}\kern\sa$^{\scriptstyle\prime}$\kern-\sb\fi}
\def\ra[#1 #2 #3.#4]{#1\sup{h}#2\sup{m}#3\sup{s}\llap.#4}
\def\dec[#1 #2 #3.#4]{#1\deg#2\arcm#3\arcs\llap.#4}
\def\deco[#1 #2 #3]{#1\deg#2\arcm#3\arcs}
\def\rra[#1 #2]{#1\sup{h}#2\sup{m}}

\def\dots{\relax\ifmmode \ldots\else $\ldots$\fi}
\def\WHzsr{\ifmmode $W\,Hz\mo\,sr\mo$\else W\,Hz\mo\,sr\mo\fi}
\def\mHz{\ifmmode $\,mHz$\else \,mHz\fi}
\def\GHz{\ifmmode $\,GHz$\else \,GHz\fi}
\def\mKs{\ifmmode $\,mK\,s$^{1/2}\else \,mK\,s$^{1/2}$\fi}
\def\muKs{\ifmmode \,\mu$K\,s$^{1/2}\else \,$\mu$K\,s$^{1/2}$\fi}
\def\muKRJs{\ifmmode \,\mu$K$_{\rm RJ}$\,s$^{1/2}\else \,$\mu$K$_{\rm RJ}$\,s$^{1/2}$\fi}
\def\muKHz{\ifmmode \,\mu$K\,Hz$^{-1/2}\else \,$\mu$K\,Hz$^{-1/2}$\fi}
\def\MJysr{\ifmmode \,$MJy\,sr\mo$\else \,MJy\,sr\mo\fi}
\def\MJysrmK{\ifmmode \,$MJy\,sr\mo$\,mK$_{\rm CMB}\mo\else \,MJy\,sr\mo\,mK$_{\rm CMB}\mo$\fi}
\def\microns{\ifmmode \,\mu$m$\else \,$\mu$m\fi}

\def\muK{\ifmmode \,\mu$K$\else \,$\mu$\hbox{K}\fi}
\def\microK{\ifmmode \,\mu$K$\else \,$\mu$\hbox{K}\fi}
\def\muW{\ifmmode \,\mu$W$\else \,$\mu$\hbox{W}\fi}
\def\kms{\ifmmode $\,km\,s$^{-1}\else \,km\,s$^{-1}$\fi}
\def\kmsMpc{\ifmmode $\,\kms\,Mpc\mo$\else \,\kms\,Mpc\mo\fi}
\providecommand{\sorthelp}[1]{}

\def\reff@jnl#1{{\rm#1\/}}

\def\aj{\reff@jnl{AJ}}                  
\def\araa{\reff@jnl{ARA\&A}}            
\def\apj{\reff@jnl{ApJ}}                
\def\apjl{\reff@jnl{ApJ}}               
\def\apjs{\reff@jnl{ApJS}}              
\def\ao{\reff@jnl{Appl.Optics}}         
\def\apss{\reff@jnl{Ap\&SS}}            
\def\aap{\reff@jnl{A\&A}}               
\def\aapr{\reff@jnl{A\&A~Rev.}}         
\def\aaps{\reff@jnl{A\&AS}}             
\def\azh{\reff@jnl{AZh}}                        
\def\baas{\reff@jnl{BAAS}}              
\def\jcap{\reff@jnl{JCAP}}              
\def\jrasc{\reff@jnl{JRASC}}            
\def\memras{\reff@jnl{MmRAS}}           
\def\mnras{\reff@jnl{MNRAS}}            
\def\pra{\reff@jnl{Phys. Rev. A}}         
\def\prb{\reff@jnl{Phys. Rev. B}}         
\def\prc{\reff@jnl{Phys. Rev. C}}         
\def\prd{\reff@jnl{Phys. Rev. D}}         
\def\prl{\reff@jnl{Phys. Rev. Lett}}      
\def\pasp{\reff@jnl{PASP}}              
\def\pasj{\reff@jnl{PASJ}}              
\def\qjras{\reff@jnl{QJRAS}}            
\def\skytel{\reff@jnl{S\&T}}            
\def\solphys{\reff@jnl{Solar~Phys.}}    
\def\sovast{\reff@jnl{Soviet~Ast.}}     
 \def\ssr{\reff@jnl{Space~Sci.Rev.}}    
\def\zap{\reff@jnl{ZAp}}                
\def\nat{\reff@jnl{Nature}}             
\def\physrep{\reff@jnl{Phys. Rep.}}      

\def\Planck{\textit{Planck}}

\def\ben{\begin{enumerate}}
\def\een{\end{enumerate}}
\def\bi{\begin{itemize}}
\def\ei{\end{itemize}}
\def\be{\begin{equation}}
\def\ee{\end{equation}}
\def\bea{\begin{eqnarray}}
\def\eea{\end{eqnarray}}
\def\ba{\begin{align}}
\def\ea{\end{align}}

\def\bdd{\boldsymbol{d }}

\def\bdm{\boldsymbol{m }}
\def\bdn{\boldsymbol{n }}

\def\bds{\boldsymbol{s }}

\makeatletter
\newcommand\footnoteref[1]{\protected@xdef\@thefnmark{\ref{#1}}\@footnotemark}
\makeatother

\title[Separating tensor, lensing, and foreground B-modes]{Joint Bayesian estimation of tensor and lensing B-modes in the power spectrum of CMB polarization data}
\author[M.\,Remazeilles et al.]{M.\,Remazeilles,$\!^{1}$\thanks{E-mail:~\url{mathieu.remazeilles@manchester.ac.uk}} C.\,Dickinson,$\!^{1}$\thanks{E-mail:~\url{clive.dickinson@manchester.ac.uk}} H.\,K.\,Eriksen,$\!^{2}$ I.\,K.\,Wehus~$\!^{2}$   \\
$^1$Jodrell Bank Centre for Astrophysics, Alan Turing Building, School of Physics and Astronomy, The University of Manchester, \\
Oxford Road, Manchester, M13 9PL, U.K.\\
$^2$Institute of Theoretical Astrophysics, University of Oslo, P.O. Box 1029 Blindern, NO-0315 Oslo, Norway\\
}

\begin{document}

\maketitle

\begin{abstract}
We investigate the performance of a simple Bayesian fitting approach to correct the cosmic microwave background (CMB) B-mode polarization for gravitational lensing effects in the recovered probability distribution of the tensor-to-scalar ratio. 
We perform a two-dimensional power spectrum fit of the amplitude of the primordial B-modes (tensor-to-scalar ratio, $r$) and the amplitude of the lensing B-modes (parameter $A_{lens}$), jointly with the estimation of the astrophysical foregrounds including both synchrotron and thermal dust emissions. Using this Bayesian framework, we forecast the ability of the proposed CMB space mission \emph{LiteBIRD} to constrain $r$ in the presence of realistic lensing and foreground contributions.
We compute the joint posterior distribution of $r$ and $A_{lens}$, which we improve by adopting a prior on $A_{lens}$ taken from the South Pole Telescope (SPT) measurement. As it applies to the power spectrum, this approach cannot mitigate the uncertainty on $r$ that is due to E-mode cosmic variance transferred to B-modes by lensing, unlike standard delensing techniques that are performed on maps. However, the method allows to correct for the bias on $r$ induced by lensing, at the expense of a larger uncertainty due to the increased volume of the parameter space. We quantify, for different values of the tensor-to-scalar ratio, the trade-off between bias correction and increase of uncertainty on $r$. 
For \emph{LiteBIRD} simulations, which include foregrounds and lensing contamination, we find that correcting the foreground-cleaned CMB B-mode power spectrum for the lensing bias, not the lensing cosmic variance, still guarantees a $3\sigma$ detection of $r=5\times 10^{-3}$. The significance of the detection is increased to $6\sigma$ when the current SPT prior on $A_{lens}$ is adopted.
\end{abstract}

\begin{keywords}
cosmic microwave background -- inflation -- early universe -- gravitational lensing: weak -- polarization -- methods: analytical
\end{keywords}


\section{Introduction}
\label{sec:intro}

On large angular scales ($\gtrsim 2^\circ$) over the sky, the amplitude of the power spectrum of the primordial cosmic microwave background (CMB) B-mode polarization, also known as tensor-to-scalar ratio $r$, if detected will provide a direct measure of the primordial gravitational waves and the energy scale of inflation \citep{Kamionkowski1997,Seljak1997,knox2002}. The detection of the large-scale CMB B-mode is challenging for many reasons. First, the cosmological signal is extremely faint, with typical r.m.s fluctuations $< 0.1$\,$\mu$K, for $r < 10^{-2}$, and optical depth to reionization, $\tau = 0.06$. 
Second, it is scrambled by strongly polarized Galactic foregrounds. Third, instrumental systematics (e.g., detector bandpass mismatch) can create spurious B-modes. Finally, gravitational lensing effects by large-scale structures transform CMB E-modes into spurious CMB B-modes. 

A new generation of CMB space missions, \emph{LiteBIRD} \citep{Matsumura2013}, \emph{CORE}\citep{ECO_mission}, \emph{PIXIE} \citep{Pixie2011}, are now being proposed to face the challenge of detecting the primordial CMB B-mode at a level of $r \lesssim 10^{-3}$ on large angular scales. Space missions are the only CMB experiments probing the full sky and thus capable of detecting primordial B-modes on reionization scales $2 \leq \ell \leq 12$, i.e., on very large angular scales ($20^\circ-90^\circ$), as long as the foreground contamination is controlled with the desired accuracy. 
The problem of foregrounds in the context of dedicated CMB B-mode space missions has been addressed in the literature over recent years \citep{Baccigalupi2004,Dunkley2009,Betoule2009,Bonaldi2011,Katayama2011,Armitage-Caplan2012,Remazeilles2016,Remazeilles2017,Errard2016,hervias_2017}. For a comprehensive review of the foregrounds and the component separation problem for B-modes, we refer to \cite{Remazeilles2017}, where different component separation techniques have been applied to dedicated simulations of the proposed CMB space mission \emph{CORE}.

Gravitational lensing introduces spurious cosmic variance from the lens-induced B-modes (hereafter, lensing B-modes), but also can bias the inference of the amplitude, $r$, of the primordial CMB B-mode power spectrum \citep[e.g.,][]{Lewis2006}. Subtracting the lensing contribution to B-modes is often referred as ``delensing'' in the literature. Much effort is being carried out to develop delensing procedures \emph{on the maps} in order to minimise the spurious B-mode cosmic variance induced by lensing \citep[e.g.,][]{Seljak2004,Simard2015}. While the lensing bias can be subtracted from the measured CMB B-mode power spectrum, the only way to correct for the lensing cosmic variance is indeed to subtract the lensing contribution directly from the CMB B-mode map. 

The lensing B-mode template that is subtracted to the CMB B-mode map is a convolution of the CMB E-mode map and a tracer of the dark matter mass distribution integrated along the line-of-sight. The mass tracer can be obtained \emph{internally} from the CMB temperature/E-mode anisotropies by mapping the CMB lensing potential through quadratic estimators \citep{Zaldarriaga1998,Hu2002}. This approach requires having both high sensitivity and high angular resolution for the CMB experiment since the small-scale anisotropies of the lensing field are the main contributors to the large-scale lensing B-mode anisotropies. For the high-resolution \emph{CORE} experiment, it has been shown that the uncertainty on B-modes due to lensing cosmic variance could be reduced by 60\% through internal delensing \citep{ECO_lensing}. For low-resolution CMB B-mode experiments like \emph{LiteBIRD}, the mass tracer can alternately be obtained from \emph{external} datasets of large-scale structures \citep{Smith2012}, such as cosmic shear maps from optical galaxy surveys \citep{Marian2007}, SKA continuum maps of 21\,cm emission \citep{Namikawa2016}, or maps of cosmic infrared background (CIB) anisotropies \citep{Sherwin2015}. The first demonstration of successful delensing of CMB temperature anisotropies with the CIB was performed on \emph{Planck} data by \cite{Larsen2016}, while for B-modes 28\% delensing has been achieved on the SPT data through the use of external CIB data \citep{Manzotti2017}.

External delensing requires that the mass tracer has significant redshift overlap and correlation with the CMB lensing field. In \cite{Planck_2013_XVIII}, it has been shown that the CIB is 80\% correlated with the CMB lensing field on small fractions of the sky. However, for external delensing of B-modes on very large angular scales, it also requires to have full-sky CIB maps that do not suffer from significant foreground contamination \citep[e.g., from][]{Planck_PIP_XLVIII,Yu2017}. 

For internal delensing, quadratic estimators of the lensing potential are non-local over the sky since their kernel has support in harmonic space, so that in principle the whole sky has to be analyzed simultaneously. Therefore, quadratic estimators must rely on the assumption that the noise distribution in the CMB map is uniform, otherwise must correct for the resulting 'mean-field' bias \citep{Hanson2009}.  
There are several sources of homogeneous noise in CMB maps:
non-uniform scanning strategy of the instruments, non-uniform morphology of the foreground residuals after component separation, numerous sky cuts, fake CMB fluctuations resulting from masked/missing data restoration processes, such as inpainting. 
Localizing the quadratic estimators in both harmonic space and pixel space, as was explored by \cite{Bucher2012}, may help in facing the issue of non-uniform noise, although it was shown to be at the expense of a variance increase for the lensing estimate. Still, mean-field corrected quadratic lensing estimators have been successfully applied by \cite{Carron2017} for the internal delensing of \Planck\ temperature data.

One more general issue with map-based delensing is the contamination of the dark matter mass templates (CMB lensing potential and CIB maps) by residual Galactic foregrounds. Residual foregrounds may induce spurious correlations between the lensing mass tracer and the CMB map that must be delensed. Moreover, lensing field reconstruction relies on exploiting the small deviations to the Gaussian statistics of the CMB, therefore non-Gaussian foreground residuals in the maps may also bias the delensing \citep{Sehgal2016,Namikawa2017}.

In this paper, we address the question of delensing CMB B-modes in the power spectrum domain within a simple Bayesian framework that allows for seamless and joint estimation of cosmological parameters and astrophysical foregrounds.
The question we address is: \emph{can we omit to subtract the lensing cosmic variance contribution to B-modes and just correct for the lensing bias in the power spectrum, nevertheless guarantee a detection of the tensor-to-scalar ratio?} In order to provide quantitative and fair results, we include foregrounds in sky simulations of the CMB satellite project \emph{LiteBIRD}, and we achieve component separation before performing the separation of the tensor and lensing contributions to the CMB B-mode power spectrum and distribution of $r$. The approach followed in this work is to correct for the lensing bias in the foreground-cleaned CMB B-mode power spectrum by fitting simultaneously the amplitude of the primordial/tensor B-mode power spectrum (parameter $r$) and the amplitude of the lensing B-mode power spectrum (parameter $A_{lens}$) in a Bayesian framework. The method is reliable because the noise-like shape of the lensing B-mode power spectrum, as well as the shape of the tensor B-mode power spectrum, are both theoretically known. Only the amplitudes are fitted for. Although such a power spectrum approach cannot remove the cosmic variance on B-modes induced by lensing, it can still remove the lensing bias on the posterior distribution of $r$. Our method can at least provide a useful cross-check for near-term B-mode experiments, aiming at $r \lesssim 10^{-2}$. In this paper, we quantify the ratio $\sigma(r) / r$, i.e the increase of variance versus the subtraction of the bias, that we obtain by ``delensing'' the foreground-cleaned CMB B-mode power spectrum.

The paper is organised as follows. In Sect. \ref{sec:bayes}, we describe the Bayesian framework to first perform foregrounds cleaning (Sect. \ref{subsec:compsep}), then debias the tensor-to-scalar ratio estimate from lensing effects (Sect. \ref{subsec:lnl}). The sky simulations for the CMB experiment \emph{LiteBIRD}, including foregrounds and lensing contamination, are briefly described in Sect. \ref{sec:simulations}. In Sect. \ref{sec:results}, we present our results for different values of $r$ by quantifying the fractional error, $\sigma(r) / r$, of such a simple delensing approach. We conclude in Sect. \ref{sec:conc}.


\section{Bayesian framework}
\label{sec:bayes}

We adopt a Bayesian framework to perform both foreground cleaning and correction for lensing effects in a self-consistent way. The Bayesian framework allows for the full propagation of CMB and foreground uncertainties to the final fitted parameters.

\subsection{Foreground cleaning with {\tt Commander}}
\label{subsec:compsep}

The {\tt Commander} algorithm \citep{Eriksen2008} is a Bayesian parametric fitting that allows to separate the different components of emission in the sky, i.e., CMB and astrophysical foregrounds. It has been successfully applied to \Planck\ data for separating the temperature anisotropies of the different sky components \citep{Planck2015_X}. More recently, it has been employed in the context of B-mode component separation in \cite{Remazeilles2016, Remazeilles2017}. 

The method consists of fitting a parametric model of the sky, $\bdm(\nu,p)$, to a set frequency data of sky observations, $\bdd(\nu,p)$, in each pixel $p$ and at each frequency $\nu$:
\bea
\label{eq:fit}
\bdm(\nu,p) &=& a(\nu)\,\bds^{cmb}(p) \cr
            &+& \left({\nu\over \nu_0^{s}}\right)^{\beta_s}\, \bds^{sync}(p) \cr
            &+& \left({\nu\over\nu_0^{d}}\right)^{\beta_d}{B_{\nu}\left(T_d\right)}\,\bds^{dust}(p) \cr
            &+& \bdn(\nu,p),
\eea
 in units of brightness temperature, where $\bds^{cmb}(p)$ is the ${Q,U}$ amplitude of the CMB polarization
anisotropies, $\bds^{sync}(p)$ is the amplitude of the polarized
synchrotron radiation, $\bds^{dust}(p)$ is the amplitude of
the polarized thermal dust radiation, $\bdn(\nu,p)$ is the
instrumental noise in the Stokes parameters, $a(\nu)$ is
the frequency spectrum of the CMB, $\beta_s$ is the synchrotron
spectral index, $\beta_d$ is the thermal dust spectral index, and
$T_d$ is the dust temperature. 

The Bayesian component separation consists of computing the joint CMB-foreground
posterior distribution for the amplitudes of CMB and foregrounds, $\bds=(\bds^{cmb}, \bds^{sync}, \bds^{dust})$, the foreground spectral indices, $\boldsymbol{\beta}=(\beta_s,\beta_d,T_d)$, and the CMB power spectra, $C_\ell=\left\{C_\ell^{EE},C_\ell^{BB}\right\}$:
\bea
\label{eq:posterior}
\mathcal{P}\left(\bds,\boldsymbol{\beta},C_\ell \big| \bdd \right) \propto \mathcal{L}\left(\bdd \big| \bds,\boldsymbol{\beta},C_\ell\right)\mathcal{P}\left(\boldsymbol{\beta}\right),
\eea 
where $\mathcal{P}\left(\boldsymbol{\beta}\right)$ are Gaussian prior distributions on the foreground spectral indices, and $\mathcal{L}$ is the likelihood of the data given the model. The full posterior distribution Eq.~\ref{eq:posterior} can be drawn by sampling the different parameters iteratively in each pixel through the Markov Chain Monte Carlo (MCMC) Gibbs sampling scheme \citep{Wandelt2004, Eriksen2004}:
\bea
\label{eq:gibbs}
\widehat{\bds}^{(i+1)} &\leftarrow& P\left(\widehat{\bds} \big| \widehat{C}_\ell^{(i)},\boldsymbol{\widehat{\beta}}^{(i)},\bdd\right),\cr
\widehat{C}_\ell^{(i+1)} &\leftarrow& P\left(\widehat{C}_\ell \big| \widehat{\bds}^{(i+1)}\right),\cr
\boldsymbol{\widehat{\beta}}^{(i+1)} &\leftarrow& P\left(\boldsymbol{\beta} \big| \widehat{\bds}^{(i+1)}, \bdd\right),
\eea 
where the conditional probability distributions involved in the Gibbs chain Eq.~\ref{eq:gibbs} have much simpler analytic forms to implement than the full posterior distribution Eq.~\ref{eq:posterior} \citep{Eriksen2008}. The Gibbs sampling chain Eq.~\ref{eq:gibbs} mathematically converges with the increased number of Gibbs iterations to the exact joint posterior distribution Eq.~\ref{eq:posterior} \citep{Wandelt2004}.

By marginalising over the amplitudes and spectral indices, {\tt Commander} allows us to draw the posterior distribution of the foreground-cleaned CMB B-mode power spectrum, $\widehat{C}_\ell^{BB}$, and thus recover the mean and uncertainties without bias.

\subsection{Delensing with Blackwell-Rao}
\label{subsec:lnl}

The {\tt Commander} component separation technique provides MCMC Gibbs samples of CMB power spectra, $\widehat{C}_\ell^{(i)}$,  as an output of component separation. Given a set of Gibbs samples of CMB power spectra, we can make use of the Blackwell-Rao approximation \citep{Chu2005} to compute the posterior distribution of the tensor-to-scalar ratio, $r$, and the amplitude of lensing, $A_{lens}$, in a self-consistent Bayesian framework.

After component separation has been achieved we estimate the cosmological parameters $r$ and $A_{lens}$ by minimising the log-likelihood
\bea
\label{eq:br1}
-2\ln \mathcal{L}\left[\widehat{C}_\ell|C_\ell^{th}\right] 
= \sum_{\ell} (2\ell+1)\left[ \ln\left({C_\ell^{th}\over\widehat{C}_\ell}\right) + {\widehat{C}_\ell\over C_\ell^{th}} - 1\right].
\eea
The theoretical CMB B-mode power spectrum $C_\ell^{th}$ is the combination of two different templates: 
\bea
\label{eq:br2}
C_\ell^{th} = \left(\frac{r}{0.1}\right)\,C_\ell^{tensor}(r=0.1)\, +\, A_{lens}\,C_\ell^{lensing}(r=0),
\eea
where $C_\ell^{tensor}(r=0.1)$ is the tensor B-mode power spectrum template, for a tensor-to-scalar ratio $r=10^{-1}$, and $C_\ell^{lensing}(r=0)$ is the lensing-induced B-mode power spectrum template ($5$\,$\mu$K.arcmin noise-like for B-modes). We vary both $r$ and $A_{lens}$, and we make use of the Blackwell-Rao approximation to estimate the joint posterior distribution of $r$ and $A_{lens}$:
\bea
\label{eq:br3}
\mathcal{P}\left(r,A_{lens}\right) \approx {1\over N}\sum_{i=1}^N \mathcal{L}\left[\widehat{C}_\ell^i|C_\ell^{th}\left(r,A_{lens}\right)\right] \mathcal{P}^{prior}\left(A_{lens}\right),
\eea 
where the sum runs over the $N$ Gibbs samples $\widehat{C}_\ell^i$ obtained after component separation. The Blackwell-Rao estimate becomes an exact approximation of the posterior distribution of $r$ and $A_{lens}$ as the number of Gibbs samples increases \citep{Chu2005}.

The Bayesian framework allows us to adopt a Gaussian prior, $\mathcal{P}^{prior}\left(A_{lens}\right)$, on the amplitude of the lensing B-mode in Eq.~\ref{eq:br3}. This amplitude has been measured with increasing precision, e.g. at $2\sigma$ by the POLARBEAR collaboration \citep{Polarbear2014}, at $4\sigma$ by the SPT collaboration \citep{SPT2015}, and at $7\sigma$ by the BICEP2/\emph{Keck Array} collaboration \citep{BK2016}. It has also been derived indirectly from minimum-variance estimates of the lensing potential by SPT \citep{Story2015} and \Planck\ \citep{Planck2015_lensing}. To be as model-independent as possible, we choose a prior based on a direct measurement of the lensing B-mode power spectrum. Although BICEP2/\emph{Keck Array} has the most stringent constraint on $A_{lens}$ at large angular scales to date, we have checked that using it as a prior only slightly increases the significance of the detection of $r=5\times 10^{-3}$ by $3$\% with respect to the significance obtained through the use of the SPT prior. We thus opt for the $4\sigma$ measurement $A_{lens} = 1.08 \pm 0.26$ by SPT \citep{SPT2015}, exploiting the fact that the SPT constraint is derived from small angular scales, and is therefore statistically independent from the multipoles we employ for our analysis.

In this work, we will compare the results on the posterior distribution of the tensor-to-scalar ratio, $\mathcal{P}(r)$, either without any prior or with the SPT prior ${A_{lens} = 1.08\pm 0.26}$ \citep{SPT2015}. 
 The success in separating the tensor and lensing B-modes relies on one hand on the known shapes of the primordial B-mode and the lensing B-mode power spectra, on the other hand on the break of degeneracy at the reionization scales ($2 \leq \ell \leq 12$) between both power spectra.

\begin{table}
 \centering
  \begin{tabular}{lll}
\hline\hline
Frequency   & Beam FWHM  & $Q$ and $U$ noise r.m.s \\
$[\rm GHz]$   & $[\rm arcmin]$  & $[\rm \mu K.arcmin]$  \\
\hline 
40 & 69 & 36.8 \\
50 & 56 & 23.6 \\
60 & 48 & 19.5 \\
68 & 43 & 15.9 \\
78 & 39 & 13.3 \\
89 & 35 & 11.5 \\
100 & 29 & 9.0 \\
119 & 25 & 7.5 \\
140 & 23 & 5.8 \\
166 & 21 & 6.3  \\
195 & 20 & 5.7 \\
235 & 19 & 7.5 \\
280 & 24 & 13. \\
337 & 20 & 19.1 \\
402 & 17  & 36.9 \\
\hline
   \end{tabular}
   \caption{Instrumental specifications for the extended version (Hazumi \& Matsumura, private communication) of the \emph{LiteBIRD} mission \citep{Matsumura2013}, adapted from \url{http://ltd16.grenoble.cnrs.fr/IMG/UserFiles/Images/09_TMatsumura_20150720_LTD_v18.pdf}.}
  \label{tab:specs}
\end{table}


\section{Simulations}
\label{sec:simulations}

By using the Planck Sky Model (PSM) software \citep{Delabrouille2013}, we perform simulations of polarized sky observations in $15$ frequency bands for the CMB space mission \emph{LiteBIRD}. Table~\ref{tab:specs} lists the instrumental specifications of the extended version of \emph{LiteBIRD}. Our simulations include CMB, Galactic foregrounds (thermal dust and synchrotron), lensing contamination, and white thermal noise in each frequency bands, whose the r.m.s is given in Table~\ref{tab:specs}.

The lensed CMB $Q$ and $U$ Stokes parameter maps are Gaussian fields that we have simulated from the lensed CMB E- and B-mode angular power spectra generated by the Boltzmann solver {\tt CAMB} \citep{Lewis2000}. For the large scales used in this analysis ($\ell < 47$, or angular scales $\gtrsim 3^\circ$), the non-Gaussianity of lensing B-mode fluctuations \citep{Smith2004} can be neglected with respect to the Gaussian fluctuations of the primordial B-modes \citep{SmithS2006}. The likelihood Eq.~\ref{eq:br1} is thus relevant at those large angular scales where the CMB B-modes fluctuations can be approximated as a Gaussian field. To strengthen this assertion, in Sect. \ref{subsec:ng} we compare the estimates of the tensor-to-scalar ratio $r=10^{-3}$ using likelihood Eq.~\ref{eq:br1} for either a Gaussian or a non-Gaussian lensed CMB B-mode, where in the latter the lensing effects are generated on the CMB map through the {\tt ilens} routine of the PSM \citep{Delabrouille2013}. We assume a ${\rm \Lambda CDM}+r$ cosmology, with tensor-to-scalar ratios ranging from $r=10^{-1}$ to $r=10^{-3}$, optical depth to reionization $\tau = 0.055$ \citep{Planck_lowl_2016}, $A_{lens}=1$, and the other cosmological parameters set to the \emph{Planck} 2015 best-fit values \citep{planck2015_overview}. The CMB component is scaled uniformly across frequencies given that its spectrum is achromatic in thermodynamic units.

The polarized Galactic synchrotron radiation is simulated by extrapolating the \emph{WMAP} 23\,GHz polarization maps \citep{Page2007,Bennett2013}, $Q_{23\, GHz}$ and $U_{23\,GHz}$, to \emph{LiteBIRD} frequencies through a power-law frequency dependence:
\bea
Q^{sync}_\nu &=& Q_{23\, GHz}\left({\nu\over 23\,{\rm GHz}}\right)^{\beta_s}, \cr
U^{sync}_\nu &=& U_{23\, GHz}\left({\nu\over 23\,{\rm GHz}}\right)^{\beta_s},
\eea
with a synchrotron spectral index $ \beta_s = -3$. The chosen value of $\beta_s$ is close to the typical mean values of the synchrotron spectral index measured at CMB frequencies in the literature \citep[e.g.,][]{Davies1996,Kogut2007,Miville-Deschenes2008,Dickinson2009,Bennett2013,Planck2015_X}.

The polarized Galactic thermal dust radiation is generated from the intensity map of the \Planck\ {\tt GNILC} dust model \citep{Planck_PIP_XLVIII} as:
\bea
Q^{dust}_\nu &=& f_d\,g_d\,I^{\tt GNILC}_\nu\,\cos\left(2\gamma_d\right), \cr
U^{dust}_\nu &=& f_d\,g_d\,I^{\tt GNILC}_\nu\,\sin\left(2\gamma_d\right),
\eea
where $I^{\tt GNILC}_\nu$ is the CIB-free {\tt GNILC} dust intensity map at the frequency $\nu$ and $f_d = 15$\,\% is the intrinsic dust polarization fraction. 
The observed polarization fraction depends on the level depolarization along the line-of-sight due to potential averaging of polarization components at different angles.  The level of depolarization is set by a geometric depolarization factor, $g_d$, which can be computed with the knowledge of the 3-D Galactic magnetic field and 3-D distribution along the line-of-sight. In our model, on large scales, the polarization angles $\gamma_d$ and the geometric depolarization factor $g_d$ for dust is essentially deduced from \emph{WMAP} 23\,GHz data for synchrotron, resulting in similar polarization angles \citep{Delabrouille2013}. We note that the exact distribution of the dust polarization angles is not important for our analysis since with {\tt Commander} we fit for synchrotron and dust (and $Q$ and $U$) independently.
The geometric depolarization lowers the average dust polarization fraction to $f_dg_d\sim 8\%$, which is similar to the mean value observed across the sky \citep{Planck2015_Int_XIX}.

The thermal dust $I$,$Q$, and $U$ maps are scaled across \emph{LiteBIRD} frequencies through a modified blackbody spectrum:
\bea
I^{\tt GNILC}_\nu = \tau^{\tt GNILC}_{353} \left({\nu\over 353\,{\rm GHz}}\right)^{\beta_d} B_{\nu}(T_d),
\eea
where the dust emissivity is $\beta_d=1.6$, the dust temperature is $T_d=19.4$\,K,  and $\tau^{\tt GNILC}_{353}$ is the \Planck\ {\tt GNILC} dust optical depth at $353$\,GHz. The chosen values of the dust spectral index and temperature correspond to the average values as measured by \cite{Planck_PIP_XLVIII} over the full sky after cleaning the Galactic dust from the CIB contamination. They are also consistent with previous full-sky estimates by \Planck\ \citep{Planck2013_XI,planck2014-XXII}. 

The B-mode power spectrum of polarized radio and infrared point-sources at CMB frequencies $\sim 100$\,GHz is expected to start dominating the primordial B-mode power spectrum on multipoles $\ell > 50$ for $r=10^{-3}$ \citep{Curto2013}.  
Since we perform B-mode component separation with {\tt Commander} on large angular scales $2 \leq \ell \leq 47$, we do not include polarized point-sources in the simulations. 

We consider simple foregrounds with uniform spectral indices in our simulations since the focus of this work is on delensing. Spectral variations over the sky must not dramatically change the results in terms of overall uncertainty, as long as the overall amplitude of the foregrounds in our simulations is correct. The results can only be biased in the case of incorrect modelling of the spectral properties during the foreground-cleaning step \citep{Remazeilles2016}. However, the {\tt Commander} algorithm provides direct goodness-of-fit by mapping the chi-square statistics, which measures the mismatch between the model and the data in each pixel. This allows to readjust the model fit a posteriori and reiterate the component separation if needed. {\tt Commander} is flexible enough to fit for complex foregrounds with non-trivial spectral properties. We refer to \cite{Remazeilles2016,Remazeilles2017} for detailed discussions on the B-mode component separation challenges with complex foregrounds. 


\section{Results}
\label{sec:results}

We now present results on three sets of simulations of increasing complexity: (i) CMB-only Monte Carlo realisations, (ii) \emph{LiteBIRD} simulations without foregrounds (i.e., CMB and noise only), and (iii) \emph{LiteBIRD} simulations with foregrounds, as described above.

\begin{figure}
  \begin{center}
    \includegraphics[width=0.5\textwidth]{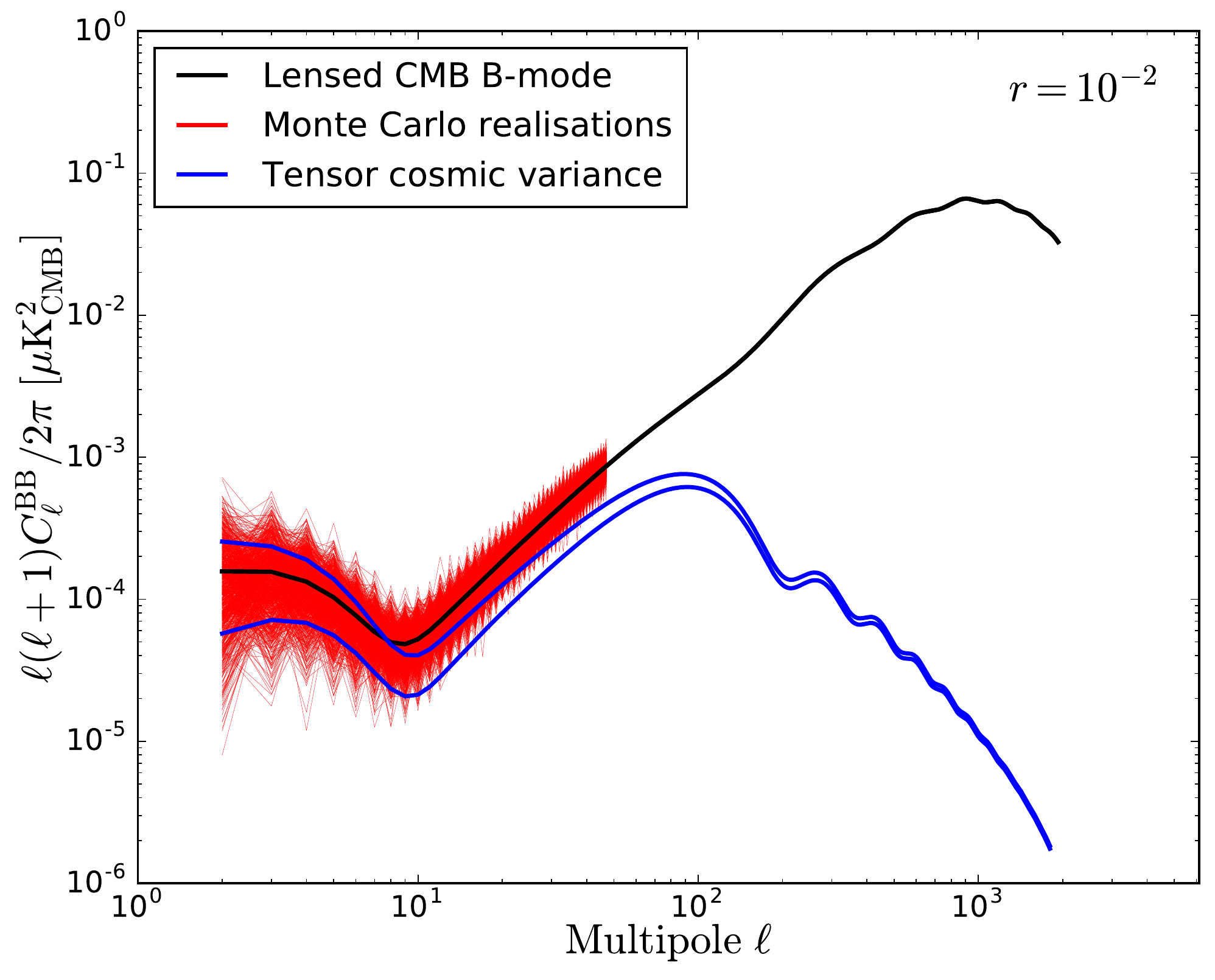}~
  \end{center}
\caption{Monte Carlo simulations of the CMB B-mode power spectrum, including both primordial and lensing B-modes, with $\tau=0.055$ and $r=10^{-2}$. \emph{Black line}: {\tt CAMB} theoretical power spectrum. \emph{Red lines}: 1000 realisations from the lensed CMB B-mode power spectrum ($\ell_{max}=47$). \emph{Blue lines}: $1\sigma$-uncertainty range allowed by the cosmic variance of tensor B-modes.}
\label{Fig:mcmc2}
\end{figure}

\begin{figure}
  \begin{center}
    \includegraphics[width=0.5\textwidth]{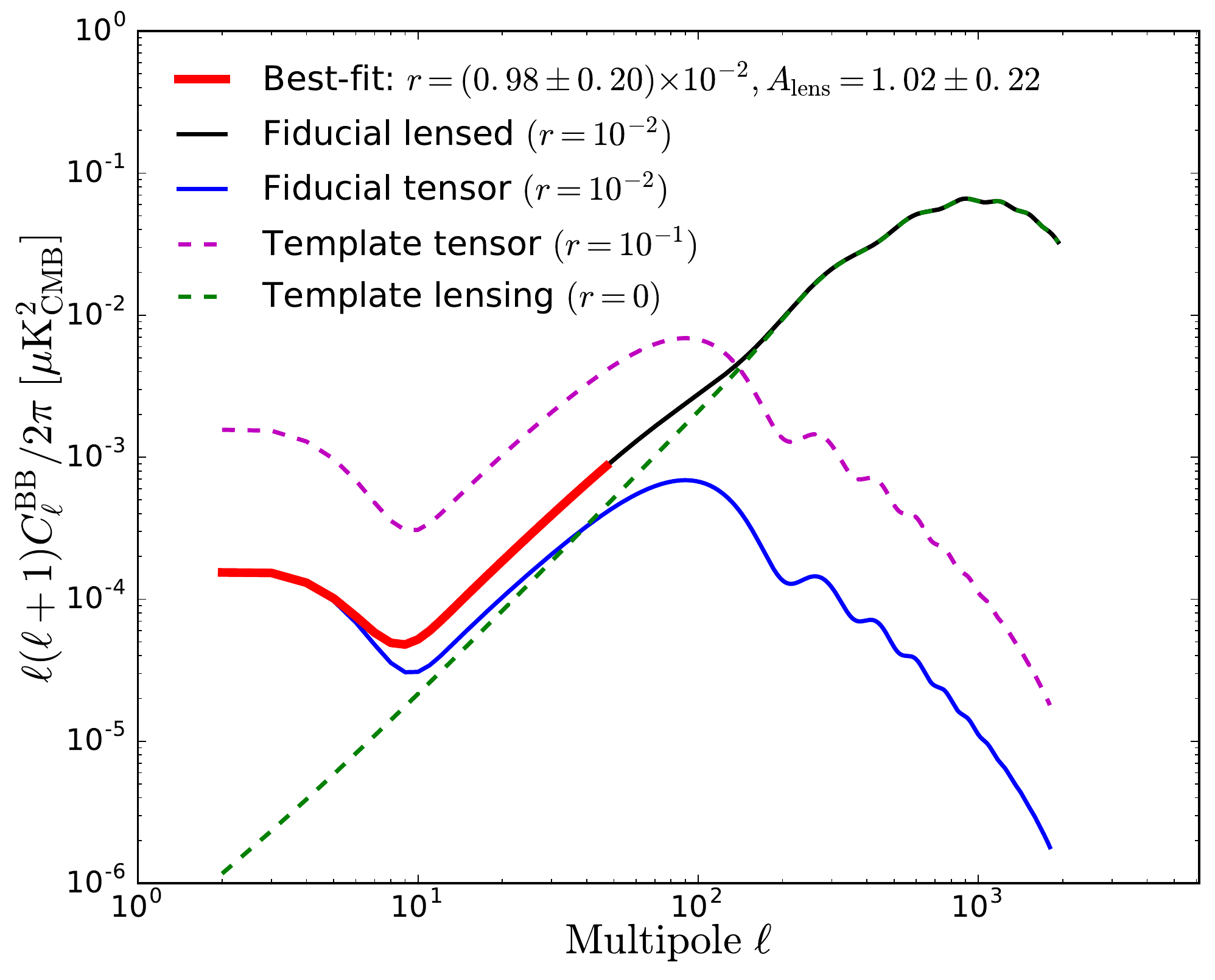}~
  \end{center}
\caption{CMB B-mode angular power spectra: input lensed CMB B-modes with $r=10^{-2}$ (\emph{black solid line}), input unlensed CMB B-modes with $r=10^{-2}$ (\emph{blue solid line}), tensor B-mode template at $r=10^{-1}$ (\emph{magenta dashed line}), lensing B-mode template (\emph{green dashed line}), maximum likelihood fit (\emph{thick red solid line}) at angular scales $2 \leq \ell \leq 47$.}
\label{Fig:ps2}
\end{figure}

\begin{figure}
  \begin{center}
    \includegraphics[width=0.5\textwidth]{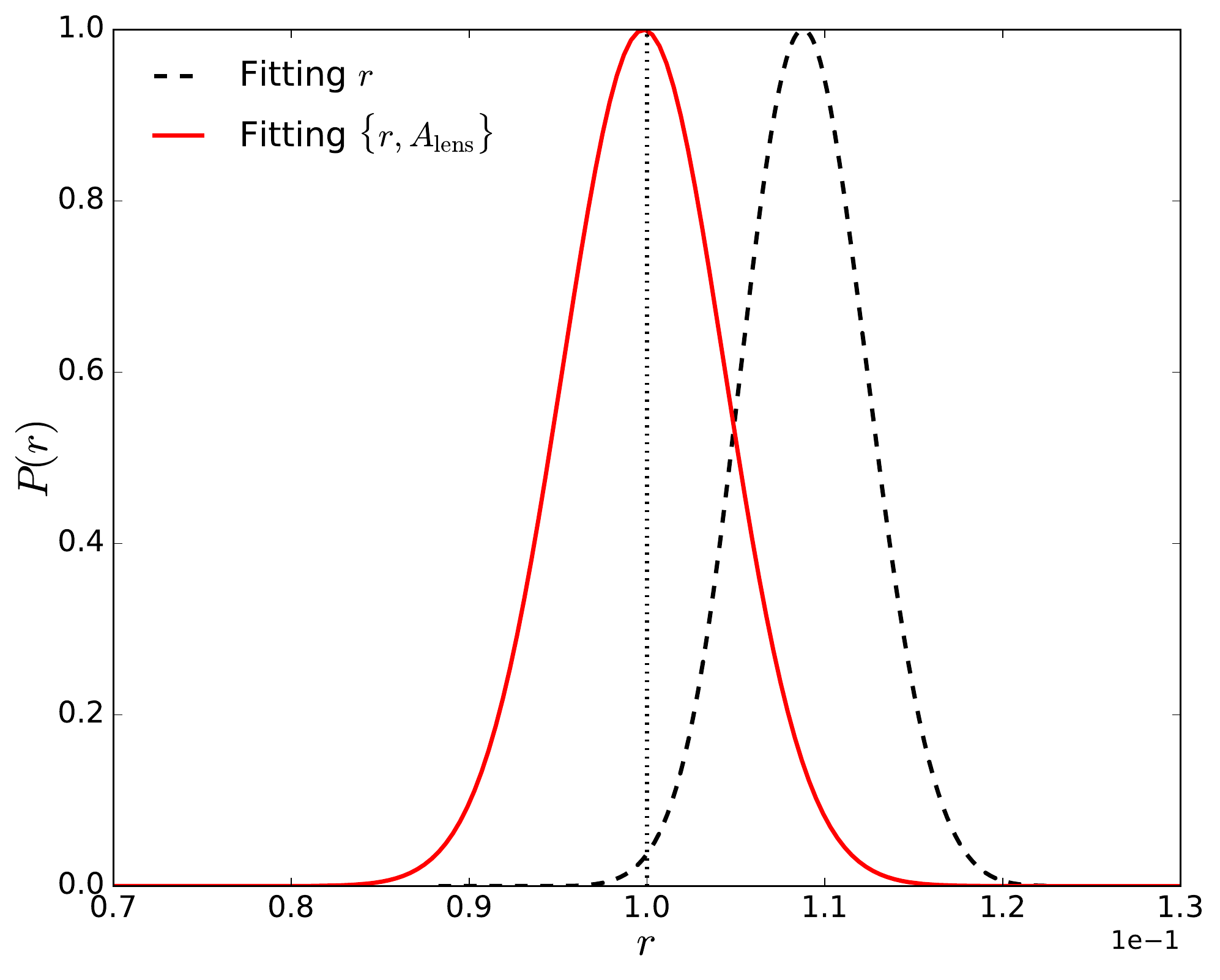}~
  \end{center}
  \caption{Recovered posterior distribution of the tensor-to-scalar ratio for the fiducial value ${r=10^{-1}}$: (i) when fitting the tensor parameter $r$ only (\emph{black dashed line}), (ii) when fitting both the tensor $r$ and lensing $A_{lens}$ parameters (\emph{red solid line}). The range of angular scales used in the likelihood is $2 \leq \ell \leq 47$.}
\label{Fig:r1}
\end{figure}

\begin{table*}
\centering
\begin{tabular}{r|ccccccc}
\hline\hline
   \multicolumn{3}{l}{\bf (a) Pure CMB Monte Carlo}   & & & & & \\ \hline
$r_{\rm true}$ & $r~[\times 10^{-p}]$ & $\underset{\mbox{w/o prior}}{\sigma(r)~[\times 10^{-p}]}$ & $\underset{\mbox{w/ prior}}{\sigma(r)~[\times 10^{-p}]}$ &$\sigma^{\rm c.v.}(r)~[\times 10^{-p}]$ & $A_{lens}$ & $\sigma\left(A_{lens}\right)$ & $\dfrac{\sigma(r)^{[\rm w/~prior]}}{\sigma(r)^{[\rm w/o~prior]}}$ \\
\hline
$10^{-1}$ & $1.00$ & $0.05$ & $0.04$  & $0.03$ & $1.01$ & $0.32$ & $0.78$ \\
$5\times 10^{-2}$ & $4.98$ & $0.33$ & $0.19$  & $0.16$ & $1.02$ & $0.31$ & $0.58$ \\
$10^{-2}$ & $0.98$ & $0.20$ & $0.07$  & $0.03$ & $1.02$ & $0.22$ & $0.34$ \\
$5\times 10^{-3}$ & $4.91$ & $1.17$ & $0.51$  & $0.17$ & $1.00$ & $0.14$ & $0.44$ \\
$10^{-3}$ & $1.00$ & $0.34$ & $0.25$  & $0.03$ & $1.00$ & $0.06$ & $0.74$ \\\hline
 &  &  &  &  &  &  & \\\hline\hline
\multicolumn{3}{l}{\bf (b) \emph{LiteBIRD}, without foregrounds}   & & & & & \\ \hline
$r_{\rm true}$ & $r~[\times 10^{-p}]$ & $\underset{\mbox{w/o prior}}{\sigma(r)~[\times 10^{-p}]}$ & $\underset{\mbox{w/ prior}}{\sigma(r)~[\times 10^{-p}]}$ &$\sigma^{\rm c.v.}(r)~[\times 10^{-p}]$ & $A_{lens}$ & $\sigma\left(A_{lens}\right)$ & $\dfrac{\sigma(r)^{[\rm w/~prior]}}{\sigma(r)^{[\rm w/o~prior]}}$ \\
\hline
$10^{-1}$ & $1.00$ & $0.05$ & $0.04$  & $0.03$ & $0.98$ & $0.32$ & $0.82$ \\
$5\times 10^{-2}$ & $4.74$ & $0.30$ & $0.20$  & $0.16$ & $1.02$ & $0.31$ & $0.65$ \\
$10^{-2}$ & $0.94$ & $0.20$ & $0.07$  & $0.03$ & $1.12$ & $0.23$ & $0.35$ \\
$5\times 10^{-3}$ & $5.86$ & $1.33$ & $0.52$  & $0.17$ & $0.87$ & $0.15$ & $0.39$ \\
$10^{-3}$ & $0.96$ & $0.32$ & $0.20$  & $0.03$ & $0.92$ & $0.05$ & $0.63$ \\\hline
 &  &  &  &  &  &  & \\\hline\hline
\multicolumn{3}{l}{\bf (c) \emph{LiteBIRD}, with foregrounds}   & & & & & \\ \hline
$r_{\rm true}$ & $r~[\times 10^{-p}]$ & $\underset{\mbox{w/o prior}}{\sigma(r)~[\times 10^{-p}]}$ & $\underset{\mbox{w/ prior}}{\sigma(r)~[\times 10^{-p}]}$ &$\sigma^{\rm c.v.}(r)~[\times 10^{-p}]$ & $A_{lens}$ & $\sigma\left(A_{lens}\right)$ & $\dfrac{\sigma(r)^{[\rm w/~prior]}}{\sigma(r)^{[\rm w/o~prior]}}$ \\
\hline
$10^{-1}$ & $0.99$ & $0.05$ & $0.05$  & $0.03$ & $0.97$ & $0.32$ & $0.87$ \\
$5\times 10^{-2}$ & $4.70$ & $0.39$ & $0.27$  & $0.16$ & $1.03$ & $0.31$ & $0.69$ \\
$10^{-2}$ & $0.94$ & $0.26$ & $0.11$  & $0.03$ & $1.02$ & $0.28$ & $0.43$ \\
$5\times 10^{-3}$ & $5.09$ & $1.60$ & $0.81$  & $0.17$ & $0.87$ & $0.20$ & $0.49$ \\
$10^{-3}$ & $1.09$ & $0.48$ & $0.44$  & $0.03$ & $0.82$ & $0.09$ & $0.92$ \\\hline
\hline
 \end{tabular}
 \caption{{\bf Results on delensing the power spectrum for (a) pure CMB Monte Carlo, (b) \emph{LiteBIRD} simulations without foregrounds, and (c) \emph{LiteBIRD} simulations with foregrounds.} \emph{First column}: true $r$ value. \emph{Second column}: reconstructed $r$ value. \emph{Third column}: 1$\sigma$-uncertainty on $r$ without any prior. \emph{Fourth column}: 1$\sigma$-uncertainty on $r$ when the SPT prior on $A_{lens}$ is used. \emph{Fifth column}: uncertainty due to cosmic variance of pure tensor modes. The values in $r$ and $\sigma(r)$ columns must be multiplied by the factor $10^{-p}$, where $p=1,2,3$ depending on the corresponding order of magnitude of $r_{\rm true}$ in each row. \emph{Sixth column}: reconstructed $A_{lens}$ value. \emph{Seventh column}: 1$\sigma$-uncertainty on $A_{lens}$. \emph{Last column}: factor of improvement on $\sigma(r)$ when applying the SPT prior on $A_{lens}$. The best-fit estimates and uncertainties on $r$ and $A_{lens}$ that are quoted in those columns are the mean and standard deviation of the joint posterior distribution drawn by the Blackwell-Rao estimator from the sample of MCMC Gibbs $C_\ell$ estimates fitted by {\tt Commander}.
}
\label{tab:summary_table}
\end{table*}

\subsection{On Monte Carlo CMB realisations}
\label{subsec:mc}

Before addressing the full joint problem including both CMB signal, foregrounds and noise, we consider the ideal problem including only CMB fluctuations in order to build intuition regarding the power spectrum estimator itself.
 We present the results of our delensing approach on $N=1000$ Monte Carlo (MC) realisations of a lensed CMB B-mode power spectrum obtained from {\tt CAMB} \citep{Lewis2000}, for different values of the tensor-to-scalar ratio ranging from $r=10^{-1}$ to $r=10^{-3}$.

In Fig.~\ref{Fig:mcmc2}, the thin red lines show the angular power spectra of $N=1000$ MC realisations of the lensed CMB B-modes for $r=10^{-2}$ and $\tau=0.055$, in the angular scale range $2 \leq \ell \leq 47$. The fiducial power spectrum from {\tt CAMB} is shown by the black line, while the $1\sigma$ interval allowed by cosmic variance of the pure tensor B-modes is drawn by the two blue lines. It should be noted that an additional cosmic variance coming from E-modes that are transformed into B-modes by gravitational lensing also contributes to the CMB B-mode realisations (red lines).

The result of fitting $r$ and $A_{lens}$ with the Blackwell-Rao estimator Eqs.~(\ref{eq:br1}), (\ref{eq:br2}), (\ref{eq:br3}) to correct for the lensing bias in the CMB B-mode angular power spectrum is shown for $r=10^{-2}$ in Fig.~\ref{Fig:ps2}. The magenta dashed line shows the tensor B-mode power spectrum template, $C_\ell^{tensor}(r=10^{-1})$, and the green dashed line the pure lensing B-mode power spectrum template, $C_\ell^{lensing}(r=0)$, that we have used in the Blackwell-Rao estimator Eq.~(\ref{eq:br3}) to fit the data provided by the 1000 MC samples of $C_\ell$s for a fiducial value of $r=10^{-2}$ (thin red solid lines of Fig.~\ref{Fig:mcmc2}). The fiducial tensor B-mode power spectrum for $r=10^{-2}$ is shown by the blue solid line, while the fiducial lensed B-mode power spectrum is shown by the black solid line. Without any prior on $A_{lens}$, we still obtain a significant and unbiased detection of $r$ and $A_{lens}$ with the Blackwell-Rao estimator, with ${r=(0.98\pm 0.20)\times 10^{-2}}$ and $A_{lens}=1.02\pm 0.22$ (also see section (a) of Table~\ref{tab:summary_table}). The result of the fit is shown by the thick red solid line in Fig.~\ref{Fig:ps2}. We are thus able to correct for the lensing bias in the distribution of the tensor-to-scalar ratio $r=10^{-2}$ directly from the angular power spectrum, while $r=10^{-2}$ is still detected at $5\sigma$ in the absence of foregrounds and noise after such a simple delensing approach.

\begin{figure}
  \begin{center}
    \includegraphics[width=\columnwidth]{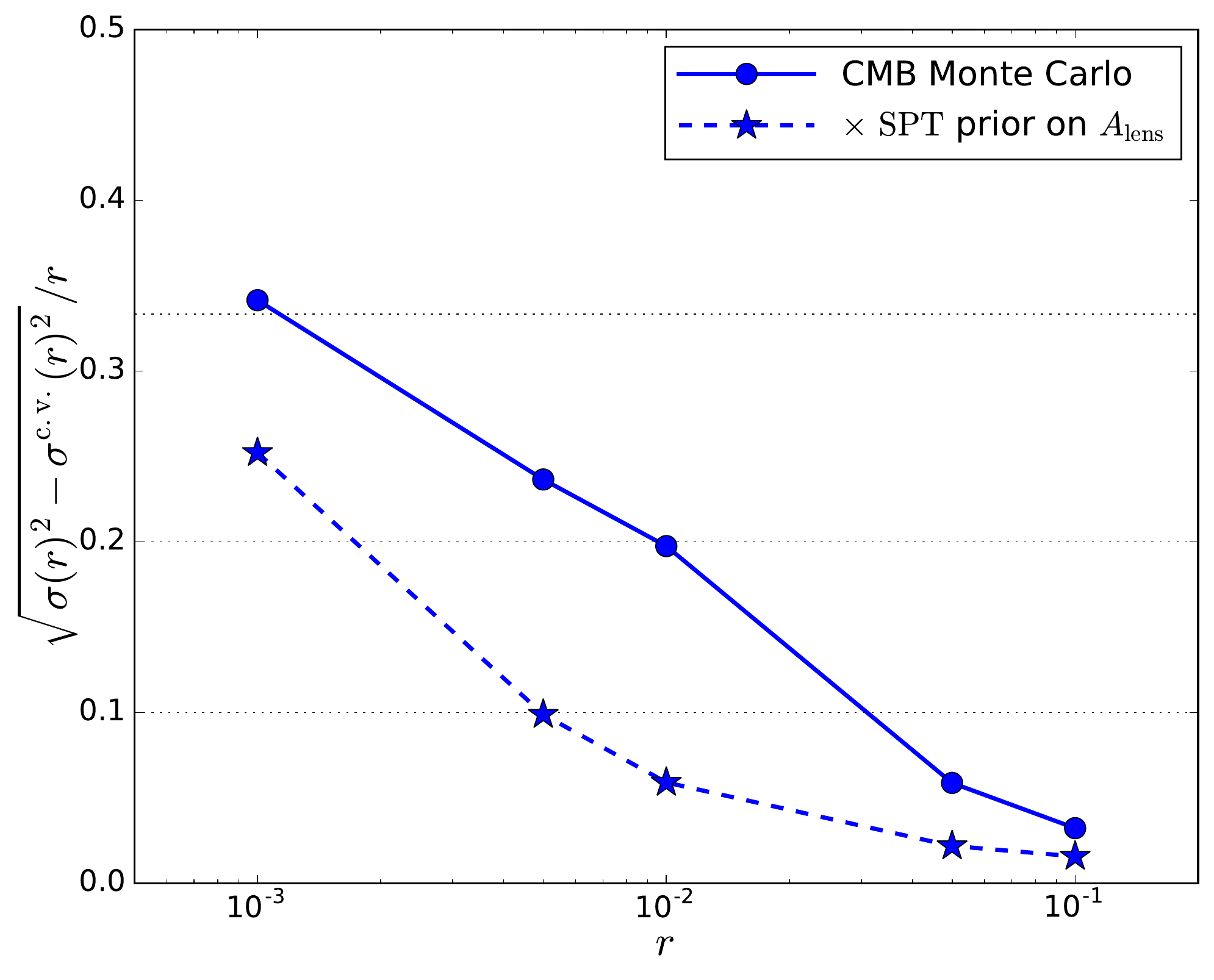}~\\
    \includegraphics[width=\columnwidth]{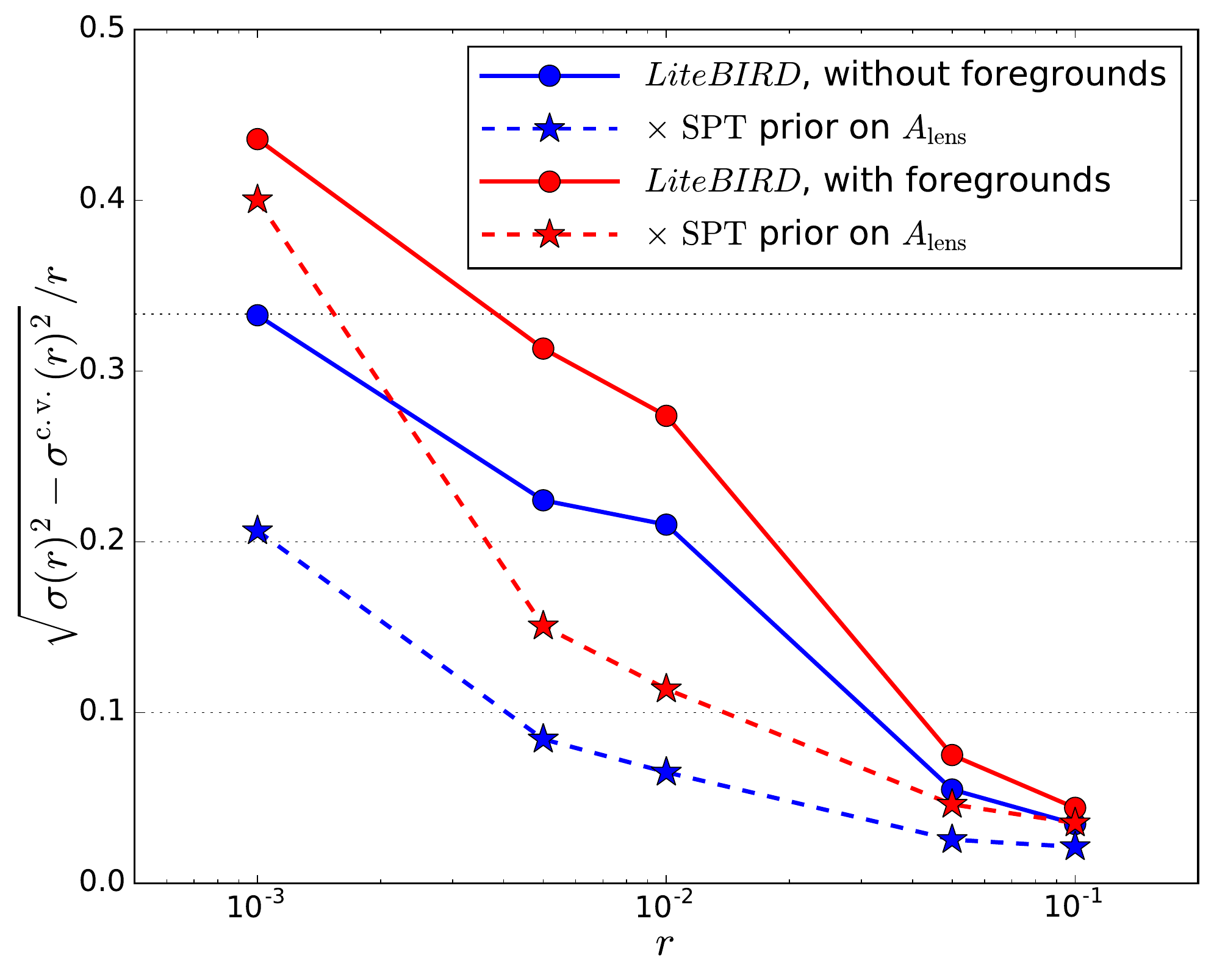}~\\
  \end{center}
\caption{Fractional error (or penalty), $\sigma(r)/ r$, versus $r$, resulting from ``delensing'' the B-mode power spectrum. \emph{Top panel}: results on pure CMB Monte Carlo, both without prior (\emph{blue solid line/circles}) and with the SPT prior on $A_{lens}$ (\emph{blue dashed lines/stars}). \emph{Bottom panel}: results on \emph{LiteBIRD} simulations, both without foregrounds (\emph{blue}) and with foregrounds (\emph{red}), and both without prior (\emph{solid lines/circles}) and with the SPT prior on $A_{lens}$ (\emph{dashed lines/stars}). The displayed fractional error actually is $\sqrt{\sigma(r)^2-\sigma^{\rm c.v.}(r)^2} / r$, so that the uncertainty due to the intrinsic cosmic variance of tensor B-modes, $\sigma^{\rm c.v.}(r)$, is removed from the overall uncertainty, $\sigma(r)$. 
The horizontal dotted lines from top to bottom show threshold limits below which the significance of the detection is more than $3\sigma$, $5\sigma$, and $10\sigma$ respectively.
}
\label{Fig:res}
\end{figure}

In Fig.~\ref{Fig:r1}, we show the result of not correcting the posterior distribution of $r$ for the lensing bias. The black dashed line shows the posterior distribution of $P(r)$ estimated by fitting only $r$ to the sample of CMB B-mode power spectra. In this case, the posterior $P(r)$ suffers from a $2\sigma$ bias due to lensing. Using our algorithm, the posterior distribution of $r$ obtained by fitting both $r$ and $A_{lens}$ on the power spectra, as shown by the red solid line, is fully debiased from the $2\sigma$ lensing bias. This simple approach still provides more than $20\sigma$ detection of $r=10^{-1}$, with best-fitting values $r=(1.00 \pm 0.05)\times 10^{-1}$ and $A_{lens}=1.01\pm 0.32$ (section (a) of Table~\ref{tab:summary_table}).

The correction for the lensing bias in the posterior distribution of the tensor-to-scalar ratio is at the cost of an increase of variance/uncertainty on $r$ because the volume of the parameter space of the likelihood has doubled. Moreover, the variance of our debiased posterior distributions, $P(r)$, still contains the spurious lensing cosmic variance from E-modes that have been transformed B-modes by lensing, something that can only be removed by a map-based delensing technique.  

 In Table~\ref{tab:summary_table} we have collected our results from the two-dimensional fit of $r$ and $A_{lens}$, for different values of the tensor-to-scalar ratio ranging from $r=10^{-1}$ to $r=10^{-3}$.
The first column displays the fiducial $r$ values, the second column collects the debiased estimates of $r$, with the associated uncertainties $\sigma(r)$ displayed in the third column when no prior is used, and in the fourth column when the SPT prior on $A_{lens}$ is adopted.
The intrinsic uncertainty on $r$ due to cosmic variance of pure tensor B-modes, $\sigma^{\rm c.v.}(r)$, is displayed in the fifth column. Sixth and seventh columns collect the estimate of $A_{lens}$ and the associated uncertainty, respectively. The factor of improvement on $\sigma(r)$ by the adoption of the prior on $A_{lens}$ is displayed in the last column.

Figure \ref{Fig:res} displays the behaviour of the fractional error, $\sigma(r)/r$, with respect to $r$, induced by ``delensing'' the power spectrum. The intrinsic and incompressible uncertainty due to cosmic variance of tensor B-modes, $\sigma^{\rm c.v.}(r)$, has been subtracted in quadrature from the overall uncertainty, $\sigma(r)$, in the actual definition of the fractional error, i.e., 
\bea
{\sigma(r)\over r} \equiv {\sqrt{\sigma(r)^2-\sigma^{\rm c.v.}(r)^2} \over r}~.
\eea
For easier reading, throughout the paper we will use the simplified notation, $\sigma(r)/r$, for the fractional error. 

The solid blue line/circles (top panel of Fig.~\ref{Fig:res}) show the results in the absence of any prior, while the dashed blue line/stars show the results when the SPT prior on $A_{lens}$ is adopted. The increase of the fractional error, $\sigma(r)/r$, is not linear with respect to $r$. In the absence of foregrounds, we see that for all values $r\geq 10^{-3}$, such a delensing procedure guarantees a more than $3\sigma$ detection of the tensor-to-scalar ratio without any bias, which increases to more than $4\sigma$ significance when the SPT prior on $A_{lens}$ is used.

\subsection{On \emph{LiteBIRD} simulations without foregrounds}

In this section, we are interested in the power-spectrum delensing forecasts that would be achieved by the \emph{LiteBIRD} experiment, under the assumption of a perfect control of the foreground contamination. Therefore, we now consider foreground-free \emph{LiteBIRD} simulations, so the data consist of noisy CMB polarization maps observed in the 15 frequency bands of the \emph{LiteBIRD} experiment.   

Unlike in Sect.~\ref{subsec:mc}, for a given value of the tensor-to-scalar ratio we now have a single realisation of the lensed CMB in the simulation, and we now apply the {\tt Commander} component separation algorithm of Sect.~\ref{subsec:compsep} to the foreground-free \emph{LiteBIRD} simulations in order to denoise the CMB polarization and produce MCMC Gibbs samples of denoised CMB B-mode power spectra. 

Like in Sect.~\ref{subsec:mc}, we then perform the correction of the posterior distribution of $r$ for the lensing bias by applying the Blackwell-Rao approach described in Sect.~\ref{subsec:lnl} to the {\tt Commander} sample of denoised CMB B-mode power spectra. 

The results of the two-dimensional fit of $r$ and $A_{lens}$ from the noisy, foreground-free, \emph{LiteBIRD} simulations are presented in the section (b) of Table~\ref{tab:summary_table}. The resulting fractional error, $\sigma(r)/r$, of our delensing approach is plotted in the bottom panel of Fig.~\ref{Fig:res}. Again, the blue solid line displays the results in the absence of any prior, while the blue dashed line shows how the fractional error is mitigated and the detection level improved when the SPT prior on $A_{lens}$ is exploited. In the absence of foregrounds, the delensing results from the \emph{LiteBIRD} simulations (section (b) of Table~\ref{tab:summary_table} and bottom panel of Fig.~\ref{Fig:res}) are compatible with the delensing results from pure CMB Monte Carlo realisations (section (a) of Table~\ref{tab:summary_table} and top panel of Fig.~\ref{Fig:res}), therefore showing that with \emph{LiteBIRD} the overall uncertainty on $r$ will no longer be limited by instrumental noise but goes down to the lensing cosmic variance limit. 

In the absence of foregrounds, the unbiased estimates of $r$ for \emph{LiteBIRD} after correction for the lensing bias still provides a $20\sigma$ (resp. $25\sigma$) detection of $r=10^{-1}$, a $5\sigma$ (resp. $14\sigma$) detection of $r=10^{-2}$, and a $3\sigma$ (resp. $5\sigma$) detection of $r=10^{-3}$ without prior on $A_{lens}$ (resp. with prior on $A_{lens}$). For $r=10^{-3}$ the use of the SPT prior on $A_{lens}$ reduces the overall uncertainty on $r$ by more than 30\%. While this appears to be better than in the case noise-free MC simulations, it should be recalled that those two cases are actually quite different exercises so that they cannot be compared on a strict equal basis. In case (b) of Table~\ref{tab:summary_table}, 15 frequency maps are passed through the {\tt Commander} algorithm for noise cleaning by fitting jointly for signal and noise fluctuations in each pixel prior to fitting for $r$ and $A_{lens}$ on the noise-cleaned power spectra. Therefore, in case (b) the joint distribution of $r$ and $A_{lens}$ is estimated from \emph{fitted} CMB power spectra, while in case (a) of Table~\ref{tab:summary_table} the joint distribution of $r$ and $A_{lens}$  was estimated from \emph{simulated} pure-CMB power spectra.

\subsection{On \emph{LiteBIRD} simulations with foregrounds}

We also are interested in the impact of foregrounds on the delensing capabilities. Component separation always leaves a non-zero amount of foreground residuals in the reconstructed CMB B-mode, which may degrade the detection forecasts on $r$.  

The section (c) of Table~\ref{tab:summary_table} displays the cumulative result of foreground removal and delensing in our Bayesian framework, in terms of the best-fit values of $r$ and $A_{lens}$. The behaviour of the corresponding fractional error is plotted as red lines in the bottom panel of Fig.~\ref{Fig:res}. We clearly see that the impact of residual foregrounds on the delensing results is negligible for $r \geq 5\times 10^{-2}$, while it becomes stronger and stronger in a non-linear way toward lower tensor-to-scalar values, e.g. by lowering the detection of $r=10^{-3}$ to less than $3\sigma$ (solid red line/red circles). The detection of $r=10^{-3}$ is slightly improved by the adoption of the SPT prior on $A_{lens}$ in the delensing estimator, now approaching $2.5\sigma$ (dashed red line/red stars). Our results are consistent with the detection forecasts on $r$ from combining the \emph{PIXIE} and \mbox{CMB-Stage IV} experiments \citep{Calabrese2017}. We also see in Fig.~\ref{Fig:res} (and in the last column of Table~\ref{tab:summary_table}) that the use of the SPT prior on $A_{lens}$ has most impact on tensor-to-scalar ratio values $r=5\times 10^{-3}$ and $r=10^{-2}$. For example, the SPT prior improves the significance of the detection of $r=5\times 10^{-3}$ for \emph{LiteBIRD} by a factor of 2, increasing from $\sim 3\sigma$ to $\sim 6\sigma$ significance in the presence of foregrounds. For larger tensor-to-scalar ratios ($r > 10^{-2}$), the impact of the $A_{lens}$ prior is negligible because the tensor and lensing power spectra are less and less degenerate. For lower tensor-to-scalar ratios ($r \leq 10^{-3}$), the impact of the $A_{lens}$ prior is also less significant because the cosmic variance of the lensing B-modes starts to dominate the cosmic variance of the primordial B-modes on a wider range of angular scales (towards low $\ell$-values). 

\subsection{About non-Gaussianity of the lensing B-mode}
\label{subsec:ng}

By the use of likelihood Eq.~\ref{eq:br1} in our analysis, we have assumed that the CMB B-mode fluctuations are Gaussian, although lensing effects must by definition introduce non-Gaussian fluctuations. However, on the large angular scales $\gtrsim 3^\circ$ ($\ell < 47$, $N_{\rm side}=16$ pixels) that are considered in this work, the lensing B-mode fluctuations must approximate as Gaussian in large $N_{\rm side}=16$ pixels, through the central limit theorem. Therefore, we have generated Gaussian CMB maps from a lensed CMB power spectrum in our set of simulations for the sake of simplicity.

In order to quantify the impact on the likelihood of non-Gaussian lensing B-modes, we generate the lensing effects directly to a {\sc HEALPix} $N_{\rm side}=2048$ CMB map through the {\tt ilens} routine of the PSM \citep{Delabrouille2013}, and degrade the pixelization to $N_{\rm side}=16$, corresponding to the smallest angular scale of the fluctuations probed in this work. As an example, we compare the results for the case of $r=10^{-3}$, the lowest tensor-to-scalar ratio value considered in our analysis. We find that the estimate from likelihood Eq.~\ref{eq:br1} shows a bias for non-Gaussian lensing B-modes, i.e. $r = (1.4 \pm 0.4)\times 10^{-3}$ in the absence of foregrounds (resp. $r = (1.5 \pm 0.5)\times 10^{-3}$ in the presence of foregrounds) , but this bias is still less than $1\sigma$. The uncertainty $\sigma(r)$ remains mostly unchanged. These results can be compared to the unbiased estimate of section (b) (resp. (c)) of Table \ref{tab:summary_table} for Gaussian CMB B-modes, where we found ${r = (0.96 \pm 0.32)\times 10^{-3}}$ (resp. $r = (1.1 \pm 0.5)\times 10^{-3}$). When the SPT prior on $A_{lens}$ is imposed to the likelihood, we find that the $1\sigma$ bias on $r$ due to non-Gaussian lensing B-modes is further reduced, with an estimate of $r=(0.91 \pm 0.15)\times 10^{-3}$ in the absence of foregrounds (resp. $r=(0.81 \pm 0.21)\times 10^{-3}$ in the presence of foregrounds). This confirms our expectations that non-Gaussian effects from lensing have no significant impact on the large angular scales considered in our analysis.

\section{Conclusions}
\label{sec:conc}

CMB polarization data from next-generation space missions, like \emph{LiteBIRD}, must allow to measure the CMB B-mode power spectrum on large angular scales over the sky after foreground cleaning. In particular, CMB B-mode space missions should enable to detect the reionization bump of the primordial B-mode power spectrum, therefore breaking the power spectrum degeneracy between primordial and lensing B-modes. Exploiting this advantage, we have implemented a self-consistent Bayesian framework allowing for both component separation and correction for lensing bias in the power spectrum. We have quantified the ability of our Bayesian method to separate the primordial and lensing B-mode contributions to the recovered distribution of the tensor-to-scalar ratio. We have provided tables of detection forecasts on $r$ for \emph{LiteBIRD} after foreground cleaning and delensing. We have considered different values of the tensor-to-scalar ratio ranging from $r=10^{-1}$ to $r=10^{-3}$.

Our simple delensing approach on the power spectrum cannot remove the lensing B-mode cosmic variance induced by E-modes, unlike map-based delensing. However, even in the presence of foreground contamination the method can still provide a $6.2\sigma$ detection of $r=5\times 10^{-3}$ for a low-resolution CMB B-mode space mission like \emph{LiteBIRD}, when the SPT prior on $A_{lens}$ is adopted in the Bayesian framework. 

The Bayesian method on the power spectrum provides a fast, simple, and complementary approach to standard delensing techniques performed on CMB maps. It can be used as an independent cross-check for the detection of primordial B-modes, especially for short-term CMB B-mode experiments since our simple delensing approach achieves significant unbiased detections for levels of $r \gtrsim 10^{-2}$.

There are several directions for future improvements of this flexible Bayesian fitting approach. A first direction is the improvement of the priors on $A_{lens}$, as well as on the foreground spectral indices, by ongoing CMB and radio ground-based experiments. Another possible improvement would be to include phase (spatial) information in our Bayesian delensing algorithm, thus allowing for removing the lensing cosmic variance, for example by incorporating recent Bayesian delensing algorithms, e.g. \textsc{LensIt} \citep{Carron_Lewis2017} or \textsc{LenseFlow} \citep{Millea2017} in our Bayesian component separation algorithm {\tt Commander}. Finally, while the component separation step with {\tt Commander} has been limited in this work to very large angular scales ($2 \leq \ell \leq 47$) due to computational cost, we should be able to extend the parametric fit to higher multipoles in the near future, therefore expecting a larger precision on $r$.    

The presence of instrumental systematics, e.g. detector bandpass mismatch and beam asymmetries, makes the detection of primordial B-modes even more challenging. However, these unknowns can in principle be incorporated in a Bayesian fitting framework like {\tt Commander}, and will be investigated in a future work.

\section*{Acknowledgements}

MR and CD acknowledge funding from the European Research Council Starting Consolidator Grant (no.~307209). CD acknowledges support from an STFC Consolidated Grant (ST/L000768/1).
We thank Anthony Challinor for helpful discussions and comments on a draft of the paper. We also thank Martin Bucher for useful discussions in the early stages of this work.
We acknowledge the use of the PSM package \mbox{\citep{Delabrouille2013}}, developed by the \emph{Planck} working group on component separation, for making the simulations used in this work. Some of the results in this paper have been derived using the {\sc HEALPix} package \citep{Gorski2005}.

\bibliographystyle{mn2e}
\bibliography{blens}


\end{document}